\documentclass{PoS}

\def\ba{\begin{eqnarray}}
\def\ea{\end{eqnarray}}
\def\be{\begin{equation}}
\def\ee{\end{equation}}

\usepackage{graphicx}

\title{Effects of lattice defects in graphene on the scattering of Charge Carriers} 

\ShortTitle{Effects of lattice defects in graphene on the scattering of charge Carriers} 

\author{\speaker{Jakson Fonseca}
\\ 
Universidade Federal de Vi\c cosa\\ 
E-mail: \email{jakson.fonseca@ufv.br}} 

\author{W.\ A.\ Moura-Melo\\ 
Universidade Federal de Vi\c cosa\\ 
E-mail: \email{winder@ufv.br}} 

\author{A.\ R.\ Pereira\\ 
Universidade Federal de Vi\c cosa\\ 
E-mail: \email{apereira@ufv.br}}

\abstract{
We study the scattering of graphene quasiparticles by topological
defects, represented by holes, pentagons and heptagons. For the case
of holes, we obtain the phase shift and found that at low
concentration they appear to be irrelevant for the electron
transport, giving a negligible contribution to the resistivity.
Whenever pentagons are introduced into the lattice and the fermionic
current is constrained to move near one of them we realize that such
a current is scattered with an angle that depends only on the number
of pentagons and on the side the current taken. Such a deviation may
be determined by means of a Young-type experiment, through the
interference pattern between the two current branches scattered by a
pentagon. In the case of a heptagon such a current is also scattered
but it diverges from the defect, preventing a interference between
two beams of current for the same heptagon.} 

\FullConference{4th International Conference on Fundamental Interactions\\ 
August 1-7, 2010\\ 
Vi\c cosa, Brazil} 

\begin{document} 

\section{Introduction and Motivation}

Condensed matter physics is a branch of physics which studies systems of many particles in the condensed, i.e. solid or liquids states. The Schr\"odinger equation is the start point of current condensed matter theory, however, in practice an generic interacting many-body system is an extremely complex system and it may not always be helpful for obtaining physical properties of an interacting many body system. The collective excitations of a many body system are collective motions of the atoms and can be viewed as particles, however, the properties of those particles can be very different from the properties of the  particles that form the many body system.

An example of this collective behavior is the graphene a flat monolayer of carbon atoms tightly packed 
into a two-dimensional ($2D$)  honeycomb lattice, 
consisting of two interpenetrating triangular sublattices  \cite{Novoselov04}.
It is the first example of a truly atomic two-dimensional $(2D)$ crystalline system and it is 
the basic building blocks for graphitic
materials such as fullerenes (graphene balled into a sphere) or carbon nanotubes 
(graphene rolled-up in cylinders) \cite{Geim07,Castro-Neto06,Geim08}.
Experimental techniques provide high-quality graphene crystallites.
up to $100 \mu m $ in size, which is sufficient for most research
purpose, including the ones considered here. In this ``perfect''
layer the charge carries can travel thousands of interatomic
distances without scattering.

Graphene provides an bridge between condensed matter physics and quantum electrodynamics because the colletive excitations are described by Dirac equation for massless particles in (2+1)D, it is a zero-gap semiconductor, in which the low
energy spectrum is correctly described by the
$(2+1)D$ Dirac-like equation for a massless particle \cite{Katsnelson06} 

\be\label{Dirac} 
i\hbar\frac{\partial}{\partial t}|\Psi\rangle =
v_F\vec{\sigma}\cdot \vec{p} |\Psi\rangle,
\ee\\
where $v_{F}$ is the Fermi velocity, which plays the role of the
speed of light ($v_{F}\approx c/300$), $\vec{\sigma}=(\sigma_{x},
\sigma_{y})$ are the $2D$ Pauli matrices, $\vec{p}=-i\hbar
\vec{\nabla}$ is the linear momentum operator and $|\Psi\rangle$ is
a two-component spinor. Therefore, the quasiparticles can be viewed
as electrons that have lost their masses or as (massless) neutrinos
that acquired the electronic charge. Such a spectrum makes graphene
a material with unique electronic properties. Its description by
means of the Dirac equation is a direct consequence of graphene's
crystal symmetry. Its honeycomb lattice is made up of two equivalent
triangular carbon sublattices $A$ and $B$, whereas its cosine-like
energy bands associated with the sublattices intersect at zero
energy $(E=0)$ near the edges of the Brillouin zone, giving rise to
a conical section spectrum
at low energies\cite{Geim07}, 
say, $|E|\,<\,1$ eV. In this honeycomb lattice, the two-component
spinor $|\Psi\rangle$ is referred to as pseudospin since it is an
index indicating two interpenetrating triangular sublattices $A$ e
$B$, which is similar to spin index (up and down) in quantum
electrodynamics (QED). It is common to regard the sublattice degree
of freedom as a pseudospin, with the $A$ sublattice being the ``up",
$|+ \rangle$ and $B$ sublattice being the ``down", $|- \rangle$. 
Since $v_{F}\ll c$, it is a slow relativistic system or a strong
coupling version of QED since the graphene's dimensionless coupling
constant, $e^{2}/\hbar v_{F}\approx 1$ much higher than its QED
analogue, the fine structure constant $e^{2}/\hbar c\approx 1/137$.
All these properties make the graphene a very
interesting system, which provides a way to probe QED phenomena, for instance, by measuring its electronic properties. Several proposals for testing some predicted, but not yet observed phenomena in QED, including the Klein paradox \cite{Katsnelson06,Katsnelson07}, 
vacuum polarization \cite{Shytov07} 
and atomic collapse \cite{Shytov+07}, 
are some topics under investigation in graphene.

Here, we would like to study the behavior of graphene quasiparticles
in the presence of defects in the crystalline structure of the
material. We shall consider three types of defects: holes, pentagons
and heptagons \cite{jakson}. All these defects can be incorporated by removing or
inserting a few carbon atoms. 
The presence of defects like pentagons (heptagons) induces positive
(negative) curvature in the material. At some extent the charge
carriers motion in the presence of pentagonal (heptagonal) defects
is identical to fermions moving in a $(2+1)D$ gravitational
space-time generated by positive (``negative'') point-like masses.
Then we may employ results from gravity to analyze some effects
concerning charge carriers in the presence of such defects.
Understanding how these defects modify
the transport properties of graphene is crucial to achieve future
electronic devices using carbon-made materials.

%
%
\section{Scattering of graphene quasiparticles by holes}


In the continuum model for the graphene, it is assumed that there is a hole of radius $r_0$ cut from the system center, located at the origin. Thus the motion of the quasiparticles is performed on a flat $2D$ support given by a non-simply connected manifold, which can be viewed as defect in the material.  This model allows for the investigation of scattering effects as a function of the hole radius and could shed some light on the high charge carrier mobilities, fact observed in graphene \cite{Katsnelson08}.

To determine the phase-shift of the scattered wave function as well as the scattering cross section one has to solve the two-dimensional Dirac equation (\ref{Dirac}) which, for the case of massless particles, can be writing in a covariant form, $i\hbar\gamma^\mu\partial_\mu\psi(x)\,=\,0,$
where the covariant derivative is $\partial_\mu = [(1/v_F)\partial/
\partial t\,, \,\partial/\partial x\,, \,\partial/ \partial y\,]$,
the $\gamma$-matrices are $\gamma^0=\sigma^3,$ $\gamma^1=i\sigma^2$
and $\gamma^3=-i\sigma^1,$ obeying
$\gamma^\mu\gamma^\nu=\eta^{\mu\nu}-i\epsilon^{\mu\nu\alpha}\gamma_\alpha$,
$\eta^{\mu\nu}$ is the Minkowski tensor metric, ${\rm
diag}(\eta^{\mu\nu})=(+1,-1,-1)$ and $\epsilon^{\mu\nu\alpha}$ is
the 3-dimensional Levi-Civita symbol $(\epsilon^{012}\equiv+1).$
(The word covariant must be used carefully because $v_F$ is not
invariant, being only a parameter and the term covariant is used to
refer only the form the equation is written.) We may expand the
solutions of the free massless Dirac equation in plane waves, once
rotational invariance allows to separate the $\theta$ variable, so
that the diagonalized angular momentum
$\mathcal{J}=-i\hbar\frac{\partial}{\partial\theta}
+\frac\hbar2\sigma^3$, yielding partial waves with angular momentum
$(n+\frac12)\hbar$, takes the form,
$\psi(\vec{r},t)=e^{i(n+\frac12-\frac12\sigma^3)\theta}u_n(r)e^{-iEt/\hbar}.$

The components of the radial spinor $u_n(r)$, given by $f_n(r)$ and
$g_n(r)$, satisfy the Bessel equation of order $n$ and $n+1$.
The solution for the radial spinor outside of the hole (i.e., for $r>r_0$) is given by:
\be\label{general solution} u_n(r)=\left(\begin{array}{c}
        f_n(r) \\
        g_n(r)
      \end{array}\right)
=\left(
          \begin{array}{c}
            B_{1n}\,J_n(k r) + B_{2n}\,N_n(k r)\\
            B_{3n}\,J_{n+1}(k r) + B_{4n}\,N_{n+1}(k r)\\
          \end{array}
        \right),
\ee\\
where $J_n$ and $N_n$ are the Bessel functions of first and second
kinds (Neumann function), respectively, $n\,=\,0\,,\pm1\,,\pm2\,,\ldots$ is the angular-momentum
number and $k=\frac{E}{\hbar v_F} > 0\,$ (we are considering only
solutions with $E>0$ that describe the electronic dynamics) and
$B_{jn}\,\,(j\,=1\,,\,2\,,\,3\,,\,4)$ are constants. These constants
must be determined by the appropriate boundary conditions specified
to completely define the problem. From the physical point of view,
the correct boundary condition is determined by the requirement of
vanishing the net energy flux into the hole, which is a region
absent of lattice degrees of freedom. Consequently, the fields must
arrange themselves in such a way that the energy flux from the
incoming modes (asymptotically behaving like $e^{-ikr}$) exactly
cancel that from the outgoing waves.
Imposing Neumann boundary condition (NBC) on the
wave-functions, $\frac{\partial u_n}{\partial r}\Big|_{r=r_0}=0$, we
obtain:
\be\label{solution with a hole} u_n(r)=\left(
          \begin{array}{c}
            B_{1n} \big [J_n(kr)-\tan(t_n(kr_0))N_n(kr)\big]\\
            B_{3n}\big[J_{n+1}(kr)-\tan(t_{n+1}(kr_0))N_{n+1}(kr)\big] \\
          \end{array}
        \right)\,
\ee\\
where $\tan[t_n(kr_0)]=\frac{J'_n(kr_0)}{N'_n(kr_0)}\,.$
The terms proportional to Bessel (Neumann) functions describe the
incident (scattered) waves. Comparing their asymptotic behavior with
those for plane waves, giving us the phase-shift $\delta_n$ of the n-th partial wave that
completely determines the fermionic scattering:
\be \delta_n^A = t_n(kr_0)\,,\qquad \delta_n^B = t_{n+1}(kr_0)\,.\ee

Now, if we consider a small concentration of point-like defects with concentration $n_{\rm def}$ angle-depedent scattering cross section, $\sigma(\theta),$ their contribution to resistivity, $\rho$, may be estimated as
$\label{resistivity with a hole}
\rho
\simeq n_{\rm def}\frac{h}{e^2kk_F}(kr_0)^6\,$ \cite{jakson}.
This means that the scattering induced by small holes (with radius around a few angstroms; some lattice spacings) at low concentration are irrelevant for the electronic transport in graphene, giving a negligible contribution to the resistivity. For the case of a potential $V(r)=V_0$ at $r<R_0$ and $V(r)=0$ at $r>R_0,$ the estimation for this type of impurity contribution to the resistivity is\cite{Katsnelson07} 
$\rho\simeq(h/4e^2)n_{\rm def}R_0^2$, giving a negligible
contribution to the resistivity when the radius of the potential
$R_0$ is of the order of interatomic distances and at a low
concentration as above.

\section{Pentagonal and heptagonal defects in graphene lattice}


Exploring the $2D$ character and flexibility of this material, our
idea is to propose a system which one or more sectors are excised
from a graphene and the remainder is joined seamlessly.
Removing a wedge from the graphene and identifying the edges a cone results \cite{jakson}.
In fact, the missed link of each carbon atom resting
at the two edges of the remaining
graphene sheet can be, in principle, covalently bounded. 
Particularly, considering the symmetry of a graphitic sheet and the Euler theorem, it can be shown that only five types of cones (incorporating one to five pentagons) can be made of a continuous graphene sheet \cite{Ge94,Krishnan97}.
The motion of the charge carriers in an ideal conical graphene is
equivalent to that of a massless Dirac particle in a gravitational
field of a static particle of mass $M$ in a $(2+1)D$ space-time.

In the case of a cone with $n_d>0$, (the value $n_d$,
$(n_d\,=\,1\,,\ldots\,,\,5)$ is related to the conical angle
$\gamma\,,$ $\sin\frac\gamma2=1-\frac{n_d}{6}\,$)
the deficit-angle induced by the conical singularity is given by $2\pi(1-n_d/6)$. 
The pentagonal defect can be presented as a pseudo-magnetic vortex at the apex of a 
graphitic cone, being the flux of the vortex related to the deficit angle of the cone (see Ref. \cite{Sitenko07}).
Cones with a heptagon have negative curvature and are obtained by a
insertion of a angular sector in the carbon sheet. Then if $n_d<0$,
$-n_d$ counts the number of such sectors inserted into the graphene
sheet. Our aim is, therefore, to see the influences that such a
special graphene structure could induce on quasiparticle
wavefunctions (spinors); surely, these influences may create new
perspectives in the electronic transport properties, which are
determined by the quasiparticles constrained to move on the conical
surface.


To study the scattering of the carriers in graphene by topological defects we employ 
the analogy between defects in condensed matter physics and in $(2+1)$-dimensional gravity \cite{Volovik03}
as far as possible. For example, the dynamics of charge carriers  in an ideal conical graphene is equivalent to 
that of a massless Dirac particle in a gravitational field of a static point-like mass in a 
$(2+1)D$ space-time\cite{Brown88,Souza89}. 
Specifically, we shall consider one of the simplest curved manifold,
which is associated to the Schwarzschild solution in $(2+1)$
dimensions: a space-time locally flat with global nontrivial
properties. To describe this space-time we may use embedded coordinates $r$ and $\theta$ in the three-dimensional Euclidian space which extend over the complete range, $0\leq r \leq\infty\,,\, 0\leq\theta\leq 2\pi$, and describe a cone with the constraint $z=\sqrt{(\alpha^{-2}-1)(x^2+y^2)}$, being the line element given by $ds^2=dt^2-\alpha^{-2}dr^2-r^2d\theta^2$
where $\alpha=1-4GM$ \cite{Souza89}.
The attributes of the source are coded in the global properties of
the locally flat variables. All the information lies in the
non-trivial boundary conditions, which is important for the quantum
scattering of graphene charge carriers by defects, like pentagons
and heptagons.

In the case of a topological defect in graphene, it is useful to
change the gravitational term $4GM$ by the symbol $\beta$, so that
$2\pi \beta$ (for $0<\beta<1$) gives the deficit of angle measuring
the magnitude of the removed sector whereas $-2\pi\beta$ (for
$-\infty<\beta<0$) accounts for the angle in excess associated to
the insertion of a sector. The parameter $\beta$ takes only discrete
values because of the lattice symmetry of the graphene.

Before analyzing the quantum mechanical scattering by conical defects (pentagons and heptagons) let us make a digression concerning scattering of the charge carriers in graphene as classical relativistic particles. The classical equation of motion, determined by relativistic geodesic equation for the particles in a cone reads $\ddot{x}^\mu + \Gamma^\mu_{\alpha\beta} \dot{x}^\alpha\dot{x}^\beta=0,$ where the overdot indicates differentiation with respect to any convenient affine variable $\tau$ that parametrizes the path $x^\mu(\tau)$ \cite{Souza89}.
The angle of scattering $\pm\omega$ for the motion of the  particles in a cone  can be obtained by integration of the classical equations of motion and is given by \cite{Souza89}
$\label{angle scattering classical} \pm \omega=\pm
\pi(\alpha^{-1}-1)=\pm \pi\frac{\beta}{1-\beta},
$\
where $\pm$ refers to the side the charge carriers (current)
trajectory pass around the defect. Note that the result above is valid for all values
of $\beta$ despite the sector was removed or inserted. The
scattering angle above, presented in the embedded coordinate system measures the deflection of the asymptotic
motion on cone projected onto $x-y$ plane of the embedding three
dimensional space. The result above suggests that a pentagon or
heptagon may be used for deviating the planar current in graphene.

To obtain the correct current deviations in graphene we have to solve the Dirac equation (\ref{Dirac}) defined in a cone, say \cite{Souza89}: 
$i\hbar\gamma^\mu E_a\,^\mu D_\mu\psi=0\,,
$
where $D_\mu=\partial_\mu+\frac12\omega_{\mu;ab}\sigma^{ab}$, $\sigma^{ab}=\frac14[\gamma^a,\gamma^b]$, and $E_a\,^\mu$ is the {\it dreibein} in coordinates $(t,r,\theta)$. The spin connection $\omega_{\mu;ab}=-\omega_{\mu;ba}$ may be written in three dimensions as $\omega_{\mu;ab}=\epsilon_{abc}\omega_\mu\,^c$ with $\epsilon_{abc}$ the Levi-Civita symbol as before \cite{Souza89}. 
The rotational invariance of the problem enables us to choose
positive energy solutions that are simultaneously angular momentum
eigenfunctions, with eigenvalue $(n+\frac12)\hbar$:
\be
u_n(r)e^{-iEt/\hbar}=e^{i(n+\frac12-\frac12\sigma^3)\theta}\left(
                                                              \begin{array}{c}
                                                                u_n^A(r) \\
                                                                u_n^B(r) \\
                                                              \end{array}
                                                            \right)e^{-iEt/\hbar}\,,
\ee
where $n=0,\pm 1, \pm 2,\ldots\, .$ The solutions for $E>0$ are \cite{Souza89}
$u_n^A(r)=(\epsilon_n)^n J_\nu(\kappa r),$ and
$u_n^B(r)=(\epsilon_n)^{n+1} J_\nu(\kappa r).
$
Here, $J_\nu$ is the Bessel function of order $\nu\equiv
\epsilon_n/\alpha(n+(1\mp\alpha)/2)$, $n=0,\,\pm1,\,\pm2\,,\ldots,$
$\kappa=E/\hbar v_F \alpha\,,E> 0\,,\epsilon_n=\pm1$ and the same
sign has to be chosen for the upper and lower components of the
spinor $u_n(r)$. For $0<\alpha \leq 1$ or $0< \beta \leq 1$
(remember that $\alpha=1-\beta$) we must choose $\epsilon_n = {\rm
sign}(n+(1-\alpha)/2)={\rm sign}\,n\,({\rm sign}\,0\equiv\,1)$ to
have both components regular at the origin.
The asymptotic form of the Bessel functions determines  the phase shifts 
(they are identical for the upper and lower components) and are given by \cite{Souza89}: 
\ba\label{phase shift cone}
\delta_n&=&-\epsilon_n\frac{\pi}{2\alpha}((1-\alpha)n
+(1-\alpha)/2) = -\frac{\epsilon_n}{2}\pi\frac{\beta}{1-\beta}\Big(n-\frac12\Big)\,,
\ea \be \epsilon_n = {\rm sign}(n+(1-\alpha)/2)={\rm
sign}(n+\beta/2)\,.
\ee
The phase-shifts depend only on the number of sectors removed or
inserted in the graphene sheet, accounted by $\alpha=1-\beta$. If
$-\infty<\beta<0$, we need to be careful because $\epsilon_n = \pm
1$ depending upon the value of $(n+\beta/2)$ (but the phase-shifts
remains as above) and the phase-shifts depends only on the number of
sectors (heptagons) inserted in the flat graphene sheet. In the
presence of heptagons the carriers dynamics is identical to the that
movement of the electrons in the gravitational field of a negative
mass (although not possible in gravitation, this is feasible in the
present context).

Note that the phase shift (\ref{phase shift cone}) measures the
deflection of the asymptotic motion on the cone projected onto $x-y$
plane being qualitatively identical to the classical case discussed
before. When there is a pentagon into the lattice and the fermionic
current is constrained to pass around and sufficiently close to it
such a current is scattered by the defect with an angle which
depends only on the number of sectors removed in the graphene and on
the side current passed.
After passing by the pentagon the scattered current trajectories
cross and yields an interference pattern. In the case of a heptagon,
such a current is scattered but the trajectories diverge each other.

\section{Conclusions}

We have studied the scattering of graphene quasiparticles by
topological defects like holes, pentagons and heptagons. We obtain
the phase shift of the wave-function in all cases. For the case of
holes, the main contribution concerns the $s$ scattering and even in
this case they do not change the resistivity of the sample, at least
at low concentrations (like occurs to short range potential
impurities). We realize that when the fermionic current is constrained to move
near and around of pentagons and heptagons introduced in the
lattice, it is scattered with an angle that depends only on the
number of defects and on which side the current taken. Such a
deviation may be determined by means of a Young-type experiment,
through the interference pattern between the two currents scattered
by the pentagon. In the case of a heptagon such a current is also
scattered but it diverges from the defect. In addition, graphene would provide an
appealing way to experimentally explore general relativity in two
spatial dimensions since such effects are predicted by this theory
\cite{Souza89,Burges85}.

\section{Acknowledgments}

The authors are grateful to C. Furtado and M.B. Silva-Neto for having drawn their
attention to important references and for discussion. They also
thank CNPq, FAPEMIG and CAPES (Brazilian agencies) for financial
support.

\end{document}